\documentstyle[prc,preprint,aps]{revtex}
\begin{document}
\draft
\title{Muon capture, continuum random phase approximation and
in-medium renormalization of the axial-vector coupling constant}
\author{E. Kolbe\thanks{Present address: Physics Department,
University of Basel, CH-4056 Basel, Switzerland} and K. Langanke}
\address{W. K. Kellogg Radiation Laboratory, 106-38\\
California Institute of Technology, Pasadena, California 91125 USA}
\author{P. Vogel}
\address{Physics Department\\
California Institute of Technology, Pasadena, California 91125 USA}
\date{\today}
\maketitle

\begin{abstract}
We use the continuum random phase approximation to describe the muon
capture on ${}^{12}$C, ${}^{16}$O and ${}^{40}$Ca. We reproduce the
experimental total capture rates on these nuclei to better than 10\%
using the free nucleon weak form factors and two different residual
interactions. However, the calculated rates for the same residual
interactions are significantly lower than the data if the in-medium
quenching of the axial-vector coupling constant is employed.
\end{abstract}
\pacs{PACS numbers: 24.30.Cz, 23.40.-s, 23.40.Hc}

\narrowtext
The capture of a negative muon from the atomic $1s$ orbit,
\begin{equation}
\mu^- + (Z,N) \rightarrow \nu_{\mu} + (Z-1, N+1)^*
\end{equation}
is a semileptonic weak process which has been studied for a long time
(see, e.g., the reviews by Walecka \cite{Wal} or Mukhopadhyay
\cite{Muk} and the earlier references therein). The total capture
rate has been measured for many nuclei; in some cases the partial
capture rates to specific states in the daughter nucleus have been
determined as well.

The nuclear response in muon capture is governed by the momentum
transfer which is of the order of muon mass. The energy transferred
to the nucleus is restricted from below by the mass difference of the
initial and final nuclei, and from above by the muon mass. The phase
space and the nuclear response favor lower nuclear excitation
energies, thus the nuclear states in the giant resonance region
dominate.

Since the experimental data are quite accurate, and the theoretical
techniques of evaluating the nuclear response in the relevant regime
are well developed, it is worthwhile to see to what extent the
capture rates are understood and, based on such comparison, what can
one say about possible in-medium renormalization of the various
coupling constants. In particular, there are various indications that
the axial-vector coupling constant $g_A$ in nuclear medium is reduced
from its free nucleon value of $g_A=1.25$ to the value of
$g_A\simeq1$. The evidence for such a renormalization comes primarily
from the analysis of beta decay between low-lying states of the
$(sd)$ shell nuclei \cite{Wild}. In addition, the ``missing
Gamow-Teller strength'' problem, as revealed in the interpretation of
the forward angle $(p,n)$ charge-exchange reactions \cite{pn}, is
also often quoted as evidence for quenching of $g_A$. The
Gamow-Teller strength is concentrated in the giant GT resonance at
excitation energies not very far from the energies involved in the
muon capture, although this latter process is dominated by the
transitions to the negative parity spin-dipole states.

Muon capture also depends on the induced pseudoscalar hadronic weak
current. At the free nucleon level the corresponding coupling
constant is determined by the Goldberger-Treiman relation \cite{Gold}
\begin{equation}
F_P(q^2) = \frac{2 M_p F_A (0)} {m_{\pi}^2 - q^2}\;,
\end{equation}
where $m_{\pi}$ is the pion mass and $F_A(0)\equiv g_A=1.25$. (In
muon capture one often uses a dimensionless quantity $g_P=m_{\mu}
F_P(q^2)$ at the relevant momentum transfer $q^2\simeq
-0.9m_{\mu}^2$, such that $g_P\simeq8.4$ for free protons.) In
nuclear medium $F_P$ can be again renormalized, and this
renormalization does not necessarily obey the Goldberger-Treiman
relation \cite{Magda}.

The Continuum Random Phase Approximation (RPA) has been used
successfully in the description of the nuclear response to weak and
electromagnetic probes \cite{RPA}. The method combines the usual RPA
treatment with the correct description of the continuum nucleon decay
channel. We have used this method for calculations of the muon
capture processes on ${}^{12}$C, ${}^{16}$O, and ${}^{40}$Ca. As
residual interactions we adopted the finite-range force
\cite{Nakayama} based on the Bonn potential \cite{Machleidt} and the
zero-range Landau-Migdal force with the parametrizations for
${}^{12}$C and ${}^{16}$O taken from Refs. \cite{Krewald1} and
\cite{Krewald2}, respectively. For ${}^{40}$Ca we used the standard
parametrization as, for example, given in Ref. \cite{Speth}. Note
that none of these forces have been adjusted to weak interaction data
for these nuclei. In the calculation we evaluate the capture rate at
each energy transfer $\omega$ for each multipole separately. The
momentum conservation condition is fulfilled throughout. We also use
an accurate relativistic description of the initial bound muon. Form
factors and their $q$-dependence have been adopted from Ref.
\cite{Beise}.

The results of our calculations are summarized in Figs.~1--3, which
show the capture rate as a function of excitation energy, and in
Table~I, which compares the total capture rates with data. The total
muon capture rates for ${}^{12}$C and ${}^{16}$O, as given in
Table~I, are defined as the part of the rate where the nucleus in the
final state is excited above the particle threshold and therefore
decays via particle emission. The experimental entries in Table~I
were derived by subtracting the partial muon capture rates into the
particle bound states (we used the average of the various data sets)
from the total capture rates. We would like to point out that most of
the capture rate goes to particle-unbound states. In ${}^{12}$C
capture to particle-bound levels (mainly the ${}^{12}$B ground state)
contributes about 16\% of the total rate, while in ${}^{16}$O the
bound-state contributions are roughly 10\%. As is obvious from
Table~I, our calculations reproduce the total muon capture rates into
the continuum states very well. For all three nuclei the Bonn
potential slightly overestimates the data, however, by less than
10\%. The Landau-Migdal force reproduces the continuum data for
${}^{12}$C and ${}^{16}$O remarkably well, while it underestimates
the ${}^{40}$Ca data by about 9\%.

For ${}^{12}$C and ${}^{16}$O partial muon capture rates to
particular bound levels have been also measured. We compare these
data with our calculation in Table~II. The calculated partial muon
capture rates, as given in Table~II, again reproduce the magnitude of
these rates well, with the notable exception of the transition to the
${}^{12}$B ground state. It is well known that a proper description
of this Gamow-Teller transition requires additional configuration
mixing within the $p$-shell other than provided by the ($1p$--$1h$)
RPA approach \cite{Brown}. For the capture to the $0^-$ and $1^-$
states in ${}^{16}$N we performed the calculation not only by the
continuum RPA, but also by the standard RPA which treats all states
as bound; the two methods agree with each other quite well. For the
$2^-$ state we used only the latter method since the continuum RPA
gives a much too narrow resonance in this case and the round-off
errors are too severe.

Figure 1 shows the total ${}^{12}{\rm C} (\mu^-,\nu_{\mu})
{}^{12}{\rm B}$ capture rate as a function of neutron energy $E_N$
above the $n+{}^{11}{\rm B}$ threshold in ${}^{12}{\rm B}$. Most of
this rate goes via $1^-$ and $2^-$ multipole excitations to the giant
dipole and spin-dipole resonances. This is demonstrated in Fig.~1,
which, as additional information, shows the partial contributions of
these multipoles. Giant dipole and spin-dipole excitations also
dominate the capture rates for the other two nuclei, ${}^{16}$O and
${}^{40}$Ca. For example, we find that the $1^-$ and $2^-$ multipoles
together contribute about 75\% [65\%] to the total ${}^{16}{\rm O}
(\mu^-,\nu_{\mu}) {}^{16}{\rm N}$ [${}^{40}{\rm Ca} (\mu^-,\nu_{\mu})
{}^{40}{\rm K}$] rate. Excitation of the giant quadrupole resonance
at about 20~MeV (see Fig.~1) contributes a few percent. For
${}^{40}$Ca, the $0^-$ and $3^-$ multipoles each contribute about
10\% to the rate.

In accordance with our discussion above, we find an average
excitation energy which in all cases corresponds to the regime of the
giant dipole and spin-dipole resonances in these nuclei. If we
consider the $Q$-values of the reactions, these average energies
indicate that the average neutrino energy after the capture process
is $\langle E_{\nu}\rangle\simeq 80$~MeV, while the remaining 25~MeV
of the muon mass are transferred, on average, to internal nuclear
degrees of freedom.

In Fig.~2, we compare the excitation spectrum for the ${}^{16}{\rm O}
(\mu^-,\nu_{\mu}) {}^{16}{\rm N}$ reaction as calculated for the two
residual interactions we used. Both spectra are very similar, but the
Bonn potential predicts a slightly higher excitation rate to the
giant dipole resonances, which accounts for the 10\% difference in
the total rates for the two interactions. Finally, in Fig.~3 we show
the capture rate for ${}^{40}{\rm Ca} (\mu^-,\nu_{\mu}) {}^{40}{\rm
K}$ as a function of the excitation energy in the final nucleus
${}^{40}$K, i.e., for the whole range of the nuclear response.

All results presented so far have been obtained using the {\it free}
nucleon form factors. To test the dependence of the calculated rates
we repeated our muon capture calculations, however, using
renormalized values for $g_A$ and $g_P$. We present the results in
Table~III for three models of the in-medium renormalization. In model
1 we simply quench the axial vector coupling, and keep the
Goldberger-Treiman relation intact:
\begin{equation}
\tilde{g}_p =
\frac{2 m_{\mu} M_p \tilde{g}_A} {m_{\pi}^2-q^2} \;,\;
\tilde{g}_A=1.0\;.
\end{equation}
In models 2 and 3 we modify the relation between $g_P$ and $g_A$,
using the relation $g_A = f_{\pi} g_{\pi NN}/M_p$ and two alternative
prescriptions given in Ref. \cite{Magda}. So in model 2
\widetext
\begin{equation}
\tilde{g_p} =
\frac{2 m_{\mu} f_{\pi} \tilde{g}_{\pi NN}}
{1.35 (m_{\pi}^2-q^2) -0.35 {\vec q}^{~2}} \;,\;
\tilde{g}_{\pi NN} /g_{\pi NN}
=1.0/1.25 \;,\; \tilde{g}_A=1.0\;,
\end{equation}
and in model 3
\begin{equation}
\tilde{g}_p =
\frac{0.6 m_{\mu} f_{\pi} \tilde{g}_{\pi NN}}
{m_{\pi}^2 -q^2 -0.7 {\vec q}^{~2}} \;,\;
\tilde{g}_{\pi NN} /g_{\pi NN}
=1.0/1.25 \;,\; \tilde{g}_A=1.0\;.
\end{equation}
\narrowtext

As expected from the axial vector dominance of the muon capture cross
section, the rates in all three cases displayed in Table~III are
significantly smaller than the ones obtained for the free nucleon
form factors. Depending on the adopted model for the renormalization,
the rates are lowered by 20--30\%, where the decrease is nearly the
same for all three nuclei for a given model. This decrease clearly
shows that the muon capture rate, although dominated by the axial
vector current interaction, is also affected by the vector-axial
vector interference, and by terms containing $g_P$. The induced
pseudoscalar coupling decreases the capture rate, as one can see best
in model 3 where $g_P$ is strongly reduced. Comparing the three
models of renormalization we confirm the known weak sensitivity of
the total capture rate to the variations in $g_P$. However, most
importantly, the rates obtained with the quenched form factors are in
disagreement with the data.

To test the sensitivity of our results to the adopted residual
interaction we have performed calculations for all three nuclei, in
which the overall strength of the Landau-Migdal force has been
decreased or increased by 10\%. In all of the calculations $g_A$ was
set to $g_A=1.25$. Noting that the residual interaction is repulsive
in the isovector channel and pushes the $T=1$ strength to higher
excitation energies, a weakening of the force results in the $T=1$
states, which are the only states populated in muon capture on $T=0$
targets like ${}^{12}$C, ${}^{16}$O, and ${}^{40}$Ca, residing at
lower excitation energies. Consequently, the energy of the neutrinos
$E_\nu$, which are released after the capture to these states, are
slightly higher. As the muon capture rate is proportional to
$E_\nu^{2(1+\lambda)}$ ($\lambda$ is the order of the spherical
Bessel function of the corresponding operator) \cite{Wal}, it is
increased when the residual interaction is weakened. The same
sequence of arguments shows that the rate is lowered if the
interaction is increased. These expectations are confirmed by our
calculations. For example, the total muon capture rate on ${}^{40}$Ca
is changed to $22.77 \times 10^5~{\rm s}^{-1}$ ($24.18 \times
10^5~{\rm s}^{-1}$) if the overall strength of the Landau-Migdal is
increased (decreased) by 10\%. When compared to our results, given in
Table~I, we conclude that a 10\% variation in the interaction results
in a change of the total muon capture rate of less than 3\%. The same
sensitivity is observed for the nuclei ${}^{12}$C and ${}^{16}$O,
where the same variation in the interaction changes the rate by less
than 3\% and 2\%, respectively. We conclude that the changes in the
muon capture rates induced by the in-medium renormalization of the
form factors $g_A$ and $g_P$ are noticeably larger than its
sensitivity to reasonable variations of the residual interaction.

Before drawing conclusions from our calculations, it is perhaps
worthwhile to briefly review other calculations of the muon capture
rate for the nuclei we are considering. In the classical paper of
Foldy and Walecka \cite{FV}, the authors relate the dipole capture
rate to the experimental photo-absorption cross section. In addition,
they use symmetry arguments to relate the vector and axial vector
nuclear matrix elements. In its slightly more modern version
\cite{Wal}, which uses the present free nucleon coupling constants,
the calculation gives total capture rates quite close to ours in
Table~I for ${}^{12}$C and ${}^{16}$O, and perhaps 20\% higher than
our numbers (and the experiment) for ${}^{40}$Ca.

Another calculation worth mentioning is the treatment of ${}^{16}$O
by Eramzhyan {\it et al.} \cite{Eram}. This calculation employs a
truncated shell model with ground state correlations included and
standard free nucleon coupling constants. For the bound states in
${}^{16}$N the results are similar to ours in Table~II, particularly
to our entries for the Landau-Migdal force. For the transitions to
unbound states (which the authors treat as shell model, i.e., bound
anyway) they obtain a total capture rate of $1.02 \times 10^5~{\rm
s}$, within 10\% of our result. On the other hand, Ohtsuka's
\cite{Oht} calculated capture rates in ${}^{12}$C and ${}^{16}$O are
larger than ours and the experiment by a substantial factor of 1.5
and 2.5, respectively. The author attributes this overestimate to the
neglect of ground state correlations in the adopted nuclear model. In
fact, the same calculation overestimates the photo-reaction cross
section, which has nothing to do with weak interactions or axial
vector current, by a similar factor.

All of these calculations are therefore basically compatible with our
result, and suggest that the experimental rate is best reproduced
with the free nucleon coupling constants. A similar conclusion
follows from the comparison of the measured ${}^{12}{\rm C} (\nu_e,
e^-) {}^{12}{\rm B}^*$ cross sections for the $\nu_e$ from stopped
muon decay \cite{Karmen}, with the results calculated within the same
approach as employed here \cite{CRPA}.

In contrast, the recent calculation in Ref. \cite{Oset} is performed
quite differently. It uses an approach based on the local density
approximation to the infinite nuclear matter. In Ref. \cite{Oset},
the capture rate is reduced by a factor of about two by the strong
nuclear renormalization. This renormalization seems to include both
the effects of residual interaction, which reduce the rate as we
argued above, and the effect of in-medium renormalization. It is
difficult to separate the two, and therefore difficult to compare our
results with those of Ref. \cite{Oset}.

Our calculations show that the continuum RPA method with a standard,
unadjusted residual interaction describes the muon capture rates in
the $T=0$ nuclei quite well, provided that free nucleon form factors
are used. In particular, quenching of the axial current coupling
constant suggested in the analysis of the Gamow-Teller strength would
result in a noticeable disagreement with the data. We checked that
this conclusion is not sensitive to reasonable variation in the
residual force. We stress that the muon capture is dominated by the
transitions to the negative parity dipole and spin-dipole collective
states. Our conclusion is therefore only relevant for such states.

\acknowledgements
We would like to thank Professor Magda Ericson for discussions, in
particular about the role of the pseudoscalar coupling. This work was
supported in part by the National Science Foundation, Grant No.
PHY91-15574, and by the U.S. Department of Energy, Contract
\#DE-F603-88ER-40397.

\begin{figure}
\caption{Muon capture in ${}^{12}$C as a function of the neutron
energy $E(n)$. Full line is the total rate, dashed line is for the
$1^-$ multipole, short dashed line is for the $2^-$ multipole, and
the dotted line is for the $2^+$ multipole. The calculation is for
the Landau-Migdal interaction.}
\end{figure}

\begin{figure}
\caption{Muon capture in ${}^{16}$O. Full line is the capture rate
for the Bonn potential and the dashed line is for the Landau-Migdal
force.}
\end{figure}

\begin{figure}
\caption{Muon capture rate in ${}^{40}$Ca as a function of the
excitation energy in the final nucleus ${}^{40}$K. The calculation is
for the Landau-Migdal force.}
\end{figure}

\widetext
\begin{table}
\caption[]{Comparison of calculated total muon capture rates with
experimental data \cite{Suzuki}. For ${}^{12}$C and ${}^{16}$O the
capture rates to particle-bound states have been subtracted. The
rates are given in $10^5~{\rm s}^{-1}$.}
\begin{tabular}{cd@{}ldd}
Target nucleus &
\multicolumn{2}{c}{Experiment} &
Bonn potential &
Landau-Migdal potential\\
\hline
${}^{12}$C & 0.320 & $\pm0.01$ & 0.342 & 0.334 \\
${}^{16}$O & 0.924 & $\pm0.01$ & 0.969 & 0.919 \\
${}^{40}$Ca & 25.57 & $\pm0.14$ & 26.2 & 23.4 \\
\end{tabular}
\end{table}
\narrowtext

\begin{table}
\caption[]{Partial muon capture rates for the bound states in
${}^{12}{\rm C} (\mu^-,\nu_\mu) {}^{12}{\rm B}^*$ (upper part) and
${}^{16}{\rm O} (\mu^-,\nu_\mu) {}^{16}{\rm N}^*$(lower part) in
units of $s^{-1}$, calculated for the Landau-Migdal force (LM) and
the Bonn potential (BP). In each part the theoretical results are
shown in the top two lines, followed by the measured data. Note, that
the values marked with ${}^{\#}$ were assumed to be 0, as the $0^+
\rightarrow 2^+$ transition is second-forbidden.}
\begin{tabular}{cr@{}lr@{}lr@{}lr@{}l}
Source &
\multicolumn{2}{c}{$\omega(1^+)$} &
\multicolumn{2}{c}{$\omega(1^-)$} &
\multicolumn{2}{c}{$\omega(2^-)$} &
\multicolumn{2}{c}{$\omega(2^+)$}\\
\hline
(LM) &
\multicolumn{2}{c}{25400} &
\multicolumn{2}{c}{220} &
\multicolumn{2}{c}{40} &
\multicolumn{2}{c}{$\leq 1$} \\
(BP) &
\multicolumn{2}{c}{22780} &
\multicolumn{2}{c}{745} &
\multicolumn{2}{c}{25} &
\multicolumn{2}{c}{$\leq 1$} \\
Ref. \cite{Bu70} &
6290 & $\pm 300$ &
720 & $\pm 175$ &
10 & $\pm 230$ &
\multicolumn{2}{c}{$0^{\#}$}\\
Ref. \cite{Gi81} &
& &
1080 & $\pm 125$ &
60 & $\pm 200$ &
\multicolumn{2}{c}{$0^{\#}$}\\
Ref. \cite{Mi72} &
6000 & $\pm 400$ &
890 & $\pm 100$ &
170 & $\pm 240$ &
\multicolumn{2}{c}{$0^{\#}$}\\
Ref. \cite{Mi72} &
5700 & $\pm 800$ &
700 & $\pm 400$ &
400 & $\pm 600$ &
200 & $\pm 400$\\
Ref. \cite{Ro81} &
6280 & $\pm 290$ &
380 & $\pm 100$ &
120 & $\pm 80$ &
270 & $\pm 100$\\
\end{tabular}
\begin{tabular}{cr@{}lr@{}lr@{}l}
Source &
\multicolumn{2}{c}{$\omega(0^-)$} &
\multicolumn{2}{c}{$\omega(1^-)$} &
\multicolumn{2}{c}{$\omega(2^-)$}\\
\hline
(LM) &
\multicolumn{2}{c}{1.45} &
\multicolumn{2}{c}{1.75} &
\multicolumn{2}{c}{8.10} \\
(BP) &
\multicolumn{2}{c}{1.85} &
\multicolumn{2}{c}{3.10} &
\multicolumn{2}{c}{8.65} \\
Ref. \cite{Col} &
1.1 & $\pm 0.2$ &
1.9 & $\pm 0.1$ &
6.3 & $\pm 0.7$\\
Ref. \cite{Berk} &
1.6 & $\pm 0.2$ &
1.4 & $\pm 0.2$ &
\\
Ref. \cite{Louv} &
0.85 & $\pm 0.06$ &
1.85 & $\pm 0.17$ &
\\
Ref. \cite{Will} &
1.56 & $\pm 0.17$ &
1.31 & $\pm 0.1$ &
8.2 & $\pm 1.2$\\
\end{tabular}
\end{table}

\begin{table}
\caption[]{Same as Table~I, but calculated for the three
renormalization models
of $g_A$ and $g_P$, defined in Eqs.~(3-5). The calculations have been
performed for the Landau-Migdal force. The rates are given in
$10^5~{\rm s}^{-1}$.}
\begin{tabular}{cr@{}lddd}
Target nucleus &
\multicolumn{2}{c}{Experiment} &
\multicolumn{1}{c}{Model 1} &
\multicolumn{1}{c}{Model 2} &
\multicolumn{1}{c}{Model 3}\\
\hline
${}^{12}$C &
0.320 & $\pm 0.01$ & 0.245 & 0.252 & 0.267 \\
${}^{16}$O & 0.924 & $\pm 0.01$ & 0.682 & 0.698 & 0.736\\
${}^{40}$Ca & 25.57 & $\pm 0.14$ & 17.3 & 17.8 & 18.8 \\
\end{tabular}
\end{table}


\begin{references}
\bibitem{Wal} J. D. Walecka in {\it Muon Physics II}, edited by V.W.
Hughes and C. S. Wu (Academic Press, NY, 1975) p. 113.
\bibitem{Muk} N. C. Mukhopadhyay, Phys. Rep. {\bf 30C}, 1 (1977).
\bibitem{Wild} B. H. Wildenthal, Progr. Part. Nucl. Phys. {\bf 11}, 5
(1984).
\bibitem{pn} C. D. Goodman and S. B. Bloom, in {\it Spin Excitations
in Nuclei}, edited by F. Petrovich {\it et al.} (Plenum, New York,
1983) p. 143; G. F. Bertsch and H. Esbensen, Rep. Prog. Phys. {\bf
50}, 607 (1987);
O. H\"ausser {\it et al.}, Phys. Rev. C {\bf 43}, 230 (1991).
\bibitem{Gold} M. L. Goldberger and S. B. Treiman, Phys. Rev. {\bf
111}, 354
(1958).
\bibitem{Magda} J. Delorme and M. Ericson, to be published.
\bibitem{RPA}
M. Buballa, S. Drozdz, S. Krewald, and J. Speth,
Ann. Phys. {\bf 208}, 346 (1991);
E. Kolbe, K. Langanke, S. Krewald, and F. K. Thielemann,
Nucl. Phys. {\bf A540}, 599 (1992).
\bibitem{Nakayama}
K. Nakayama, S. Drozdz, S. Krewald, and J. Speth,
Nucl. Phys. {\bf A470}, 573 (1987).
\bibitem{Machleidt}
R. Machleidt, K. Holinde, and Ch. Elster, Phys. Rep. {\bf 149}, 1
(1987).
\bibitem{Krewald1}
G. Co and S. Krewald, Nucl. Phys. {\bf A433}, 392 (1985).
\bibitem{Krewald2}
M. Buballa {\it et al.}, Nucl. Phys. {\bf A517}, 61 (1990).
\bibitem{Speth}
S. Krewald, K. Nakayama, and J. Speth, Phys. Rep. {\bf 161}, 103
(1988).
\bibitem{Beise}
E. J. Beise and R. D. McKeown, Comments Nucl. Part. Phys. {\bf 20},
105 (1991).
\bibitem{Brown}
W. T. Chou, E. K. Warburton, and B. A. Brown,
Phys. Rev. C {\bf 47}, 163 (1993).
\bibitem{FV} L. L. Foldy and J. D. Walecka,
Il Nuovo Cimento {\bf 34}, 1026 (1964).
\bibitem{Eram} R. A. Eramzhyan, M. Gmitro, R. A. Sakaev, and L. A.
Tosunjan,
Nucl. Phys. {\bf A290}, 294 (1977).
\bibitem{Oht} N. Ohtsuka, Nucl. Phys. {\bf A370}, 431 (1981).
\bibitem{Karmen} J. Kleinfellner {\it et al.}, KARMEN collaboration,
in Proceedings of the XIII International Conference
on Particles and Nuclei, Perugia, Italy, 1993, edited by
by A. Pascolini.
\bibitem{CRPA} E. Kolbe, K. Langanke, and S. Krewald,
Phys. Rev. C {\bf 49}, 1122 (1994).
\bibitem{Oset} H. C. Chiang, E. Oset, and P. Fernandez de Cordoba,
Nucl. Phys. {\bf A510}, 591 (1990).
\bibitem{Suzuki}
T. Suzuki, D. F. Measday, and J. P. Roalsvig, Phys. Rev. C {\bf 35},
2212 (1987).
\bibitem{Bu70}
Y. G. Budgashov {\it et al.}, JETP {\bf 31}, 651 (1970).
\bibitem{Gi81}
M. Giffon {\it et al.}, Phys. Rev. C {\bf 24}, 241 (1981).
\bibitem{Mi72}
G. H. Miller, M. Eckhause, F. R. Kane, P. Martin, and R. E. Welsh,
Phys. Lett. {\bf 41B}, 50 (1972).
\bibitem{Ro81}
L. Ph. Roesch {\it et al.}, Phys. Lett. {\bf 107B}, 31 (1981).
\bibitem{Col}
R. C. Cohen, S. Devons, and A. D. Canaris, Nucl. Phys. {\bf 57}, 255
(1964).
\bibitem{Berk}
A. Astbury {\it et al.}, Nuovo Cimento {\bf 33}, 1020 (1964).
\bibitem{Louv}
J. P. Deutsch {\it et al.}, Phys. Lett. {\bf 29B}, 66 (1969).
\bibitem{Will}
F. R. Kane {\it et al.}, Phys. Lett. {\bf 45B}, 292 (1973).
\end{references}
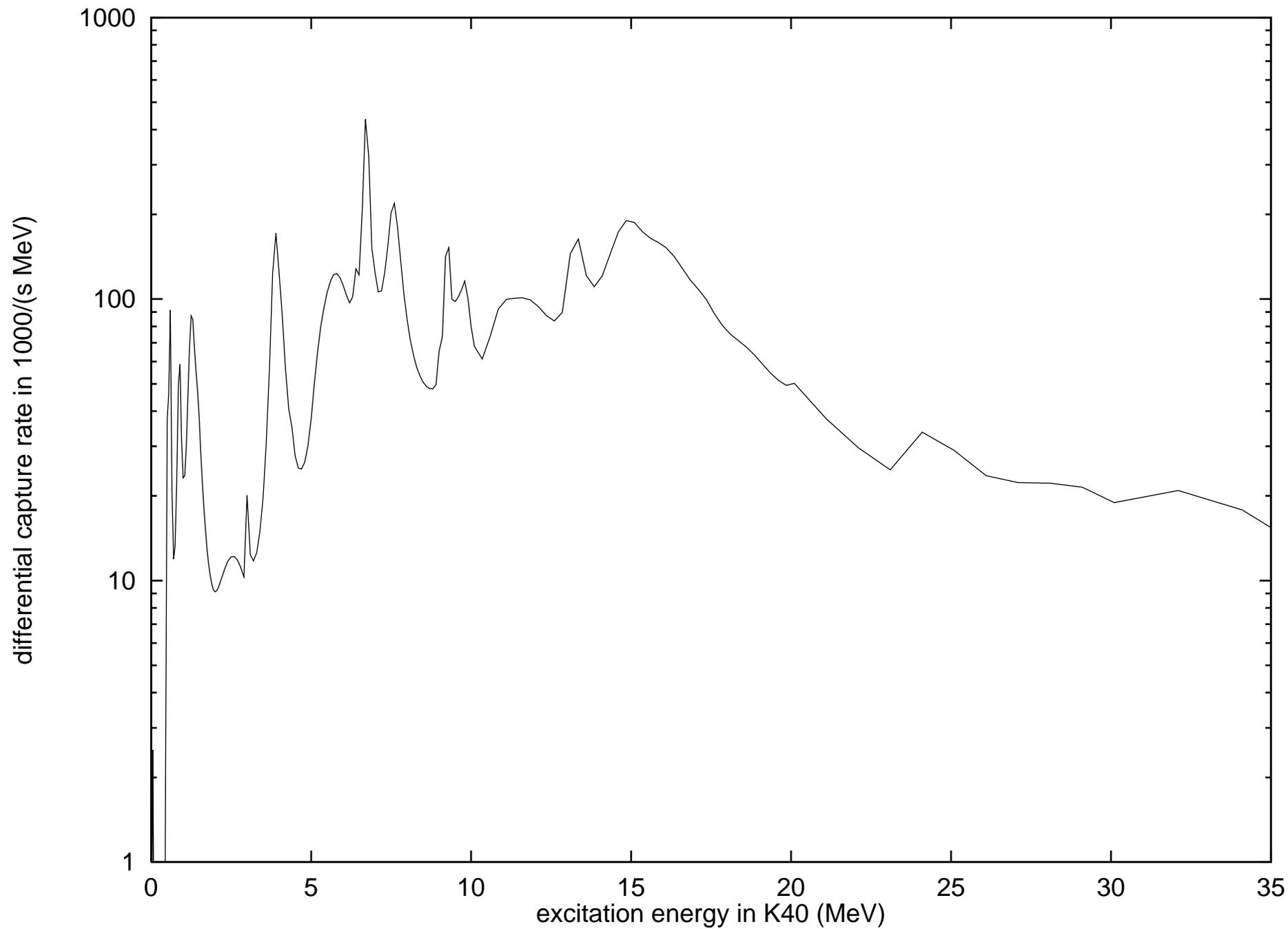
\end{document}